\documentclass[aps,pre,superscriptaddress,twocolumn]{revtex4-1}

\usepackage{setspace,tabularx} 

\usepackage{amsmath}

\usepackage{graphicx}

\newcommand{\zp}{z_{\rm up}}
\newcommand{\zm}{z_{\rm dw}}
\newcommand{\qmax}{q_{\rm max}}
\newcommand{\zpf}{\tilde{z}_{\rm up}}
\newcommand{\zmf}{\tilde{z}_{\rm dw}}
\newcommand{\h}{h}
\newcommand{\hf}{\tilde{h}}
\newcommand{\HH}{\mathcal{H}}
\newcommand{\dd}{\mathrm{d}}
\newcommand{\s}{\sigma}
\newcommand{\sadd}{\sigma_{\rm add}}
\newcommand{\sel}{\sigma_{\rm el}}
\newcommand{\st}{\tilde{\sigma}}

\begin{document}

\title{Tension moderation and fluctuation spectrum in simulated lipid membranes under an applied electric potential}

\author{Bastien Loubet}
\affiliation{MEMPHYS - Center for Biomembrane Physics, Department of Physics, Chemistry and Pharmacy, University of Southern Denmark, Campusvej 55, 5230 Odense M, Denmark}
\author{Michael Andersen Lomholt}
\affiliation{MEMPHYS - Center for Biomembrane Physics, Department of Physics, Chemistry and Pharmacy, University of Southern Denmark, Campusvej 55, 5230 Odense M, Denmark}
\author{Himanshu Khandelia}
\affiliation{MEMPHYS - Center for Biomembrane Physics, Department of Physics, Chemistry and Pharmacy, University of Southern Denmark, Campusvej 55, 5230 Odense M, Denmark}

\begin{abstract}
 We investigate the effect of an applied electric potential on the mechanics of a coarse grained POPC bilayer under tension. The size and duration of our simulations allow for a detailed and accurate study of the fluctuations. Effects on the fluctuation spectrum, tension, bending rigidity and bilayer thickness are investigated in detail. In particular the least square fitting technique is used to calculate the fluctuation spectra. The simulations confirm a recently proposed theory, that the effect of an applied electric potential on the membrane will be moderated by the elastic properties of the membrane. In agreement with the theory we find that the larger the initial tension the larger the effect of the electric potential. Application of the electric potential increases the amplitude of the long wavelength part of the spectrum and the bending rigidity is deduced from the short wavelength fluctuations. The effect of the applied electric potential on the bending rigidity is non-existent within error bars. However when the membrane is stretched there is a point where the bending rigidity is lowered due to a decrease of the thickness of the membrane. All these effects should prove important for mechanosensitive channels and biomembrane mechanics in general.
\end{abstract}

\maketitle

\section{Introduction}
The lipid bilayer that surrounds cells is an important medium as a barrier between the interior of the cells and their surroundings. In recent years, the lipid membrane of cells have been recognized to play an important role in the modulation of protein functions \cite{Lee2004,Phillips2009}. The membranes in a cell are subject to external forces and stresses that change its physical properties and in turn affect protein/membrane functions. Such change can be brought by change of the electrostatic potential that the cells maintain between the two sides of the membrane \cite{Ambjornsson2007} but it can also be due to protein activity in general \cite{Gauthier2012,Loubet2012}. Another important property of the membrane is its thermal fluctuations as these influence membrane adhesion \cite{Evans1986,Reister2008}. The properties of the fluctuations depend on the mechanical parameters of the membrane and hence it can also be modified by an applied electric potential or active proteins \cite{FarisPRL,Bouvrais2012}.

It is then of primary importance to correctly describe the effect of an additional stress on the membrane mechanical parameters with respect to a reference state.
The two main mechanical parameters of the membrane are its bending rigidity, seen as the energetic cost for inducing membrane curvature, and the tension, seen as the mechanical parameter which sets the area of the membrane.
The theoretical contribution of the applied potential to the bending rigidity and the tension have been derived in several works \cite{Ambjornsson2007,Lacoste2009}. However it has been argued in \cite{Michael2011} that the tension of a membrane subject to external stress is not the sum of the initial tension plus the additional stress. This is because the membrane is an elastic sheet that can change its area in response to the additional stress.

In this paper we investigate the effect of an applied electrostatic potential on a POPC membrane bilayer through molecular dynamics simulations. In particular we test and confirm a recently proposed theory \cite{Michael2011}, that the effect of the electric field on the membrane tension should depend on the available membrane excess area. We will also investigate the effect of an applied electric potential on the fluctuation spectrum of the membrane. From the fluctuation spectrum we will calculate the bending rigidity of the membrane with and without an applied potential. In addition we will vary the initial excess area/tension and investigate its effect on the bending rigidity.

\section{Methods}

\subsection{Theory}
We model the membrane as a smooth two-dimensional surface with height above the xy-plane $\h(x,y)$. We will model the effect of the electric field to be equivalent to adding a term proportional to the area in the free energy. In order to describe the effects of this additional tension we will follow the approach of \cite{Michael2011} and take the following Hamiltonian:
\begin{equation}
\HH = \HH_c + \HH_s + \HH_{\rm add}
\end{equation}
where $\HH_c$ is the Helfrich bending energy:
\begin{equation}
 \HH_c = \int \dd A\ \frac{\kappa}{2}\left( 2 H \right)^2
\end{equation}
here the integral is over the whole area $A$ of the membrane, $H$ is the local mean curvature (half the sum of the principal curvatures) and $\kappa$ is the bending rigidity of the membrane, i.e., its resistance to bending. $\HH_s$ is the contribution from the elastic stretching of the membrane. It reads:
\begin{equation}
 \HH_s = \frac{K_a}{2 A_0 } \left(A - A_0 \right)^2
\end{equation}
where $K_a$ is the area expansion modulus and $A_0$ is a the preferred area of the membrane. Finally $\HH_{\rm add}$ is the contribution from an additional intrinsically applied tension $\sadd$:
\begin{equation}
 \HH_{\rm add} = \s_{\rm add} A
\end{equation}
In our simulations $\sadd$ is taken to be the contribution from the applied electric potential. The membrane surface can be expanded in a series of complex exponentials:
\begin{equation}
 \h(x,y) = \sum_{q_x} \sum_{q_y} \hf(q_x,q_y) e^{i (x q_x + y q_y)}
\end{equation}
where $\hf$ are the corresponding expansion coefficients and $q_x$ and $q_y$ are wavenumbers in the $x$ and $y$ directions of the membrane, they can take the values:
\begin{equation}
 q_x = n_x \frac{2 \pi}{L_x} \quad {\rm and}\quad q_y = n_y \frac{2 \pi}{L_y}
\end{equation}
where $L_x$ and $L_y$ are the box size in the $x$ and $y$ direction and $n_x$ and $n_y$ are integers. In this approach the microscopic nature of the membrane is ignored and it is necessary to introduce a cut off for both $n_x$ and $n_y$ such that $\left| q_x \right| \leq \qmax$ and $\left| q_y \right| \leq \qmax$ where the wavelength corresponding to $\qmax$ is the wavelength at which the continuous description of the membrane breaks down. These restrictions on the values of $q_x$ and $q_y$ are implied throughout the paper.
The fluctuation spectrum is defined as the average value of the square of the absolute value of $\hf$ at equilibrium. Using standard statistical mechanics, the fluctuation spectrum can be shown to be \cite{Helfrich1973}:
\begin{equation}
\label{fluct}
 \left< | \hf(q) |^2 \right> = \frac{k_B T}{L_x L_y} \frac{1}{ \st q^2 + \kappa q^4 }
\end{equation}
where this expression depends only on the norm of the wavevector $q = \sqrt{q_x^2 +q_y^2}$. The brackets mean an ensemble average which is equivalent to a time average at equilibrium. $\st$ is an is a moderated tension that depends on  $\sadd$, $K_a$, $\kappa$ and the tension $\s_0$ of the membrane when $\sadd = 0$. According to \cite{Michael2011} $\st$ can be calculated self-consistently using the equation:
\begin{equation}
 \st - \sadd = K_a \frac{\left< A \right> - A_0}{A_0}
\end{equation}
If we want to use the projected area, $L_x L_y$, instead of $A_0$ we rewrite this equation as:
\begin{equation}
\label{snoapprox}
 \st - \sadd  = K_a u \left( \alpha(\st) + 1 - \frac{1}{u} \right) 
\end{equation}
where $u= L_x L_y/A_0$ (see the supplementary material for values of $u$ in our simulations) and $\alpha$ is the fractional excess area defined as:
\begin{equation}
\label{excessarea}
 \alpha = \frac{\left< A \right> - L_x L_y}{L_x L_y} =  \sum_{q_x} \sum_{q_y} \frac{1}{2} q^2 \left< | \hf(q) |^2 \right>
\end{equation}
The final tension is then the sum of the additional applied tension plus an elastic contribution and must be determined self-consistently. This elastic contribution will moderate the applied tension. Note that if we had assumed an equation of the form of $\HH_{\rm add}$ for $\HH_s$ instead, say $\HH_s = \s_0 A$, the tension after adding $\HH_{\rm add}$ would simply be $\st = \sadd + \s_0$ i.e. there would be no elastic response of the membrane to the external additional stress. In this paper we verify the elastic moderation of the tension through molecular dynamics simulations. In particular we will apply an electric potential difference, $V$, across our membranes. This electric potential will give rise to stress in the membrane in the form of an additional tension $\sadd$. In order to try to quantify this effect we use Eq. \ref{snoapprox} making the following assumption for $\sadd$. The membrane we simulate is symmetrical, so the first non-zero term in an expansion in $V$ of $\sadd$ is quadratic. This means that, to lowest order:
\begin{equation}
\label{anzat}
 \sadd = \mu V^2
\end{equation}
where $\mu$ is a phenomenological constant that depends on the membrane composition, but also on the ionic concentration and the water properties. In the following we will use Eq. \eqref{snoapprox} together with \eqref{anzat} in order to explain the observed behavior of our membranes.

\subsection{Simulations}
The Martini coarse grained force field \cite{Marrink2007} has been used to perform simulations of a membrane bilayer. In order to observe large wavelength fluctuations the simulated membrane was made rectangular by duplicating a equilibrated patch of membrane in the $x$ direction. This patch was then equilibrated and duplicated again until we reached the desired size. The final system consisted of 602964 particles as 4224 POPC molecules, 181884 water beads, 1200 sodium ions and 1200 chloride ions. For the water, the polarizable water model of Martini has been used \cite{Yesylevskyy2010}, note that this include a background permittivity of $2.5 \epsilon_0$. We use an ionic concentration of $\sim 90\ {\rm mM}$ which was chosen such that the membrane would be electrically screened from its periodic images (the Debye length \cite{Debye1923} is on the order of $1\ {\rm nm}$). All simulations were run at a temperature of 325 K in order to accelerate fluctuation dynamics. We performed the simulation in the ${\rm NP_zAT}$ ensemble where the pressure normal to the membrane is fixed to one bar and the projected area of the membrane is fixed in the lateral direction. This allows us to directly observe the change in the fluctuations and the tension due to the applied electric potential.

We note here that we chose the Martini force field because it gives a good trade off between performance and accuracy. Furthermore the polarizable water version of it gives a reasonable value for the dielectric permittivity of water. It is known, however, that the electrical potential in the membrane (i.e. the so called dipole electrical potential) is negative with respect to the bulk fluid in Martini simulations. Experimental evidence suggest that it is actually positive for real lipid membranes \cite{Brockman1994,Yang2008}. This problem has already been noted in the original paper for the polarizable Martini water \cite{Yesylevskyy2010}, but is not an issue here because the electrostatic stresses are proportional to the square of the electric field and do not depend on its sign.

The fluctuation spectrum is calculated and convergence is checked as explained below. Our original membrane patch was $132.18\times 10.93\times 19.04$ ${\rm nm^3}$. From this we constructed three other patches with the following scaling factor on the x axis: $0.996$, $1.02$ and $1.05$, resulting in boxes of length $131.71$, $134.82$ and $138.79\ {\rm nm}$ respectively. In the following we will refer to these different cases by their scaling factor. The stretched membrane was obtained by scaling the coordinates of all residues in the system. No stable system could be obtained by scaling the the coordinates of all residues for the shrunk membrane so we used the following method: we relaxed the constraint on the fixed size of the simulation box in the x direction while fixing the size in the z direction after creating a small vacuum above and below the membrane. The system then automatically shrank in order to conserve its volume, we obtained a box size of $131.71\times 10.93\times 19.11$ ${\rm nm^3}$ which is close to a 0.996 factor ($0.99644$). For each of these patches two initial sets of velocities were used to run two sets of simulations. They were first equilibrated with Berendsen coupling for the temperature and the pressure \cite{Berendsen1984} and then subsequently run with the Nose-Hoover \cite{Nose1984,Hoover1985} thermostat and the Parrinello-Rahman \cite{Parrinello1981,Nose1983} barostat. All electrostatics was handled using the Particle Mesh Ewald (PME) method \cite{Darden1993,Essmann1995} using the parameters for the polarizable water model \cite{Yesylevskyy2010}.
Subsequently we applied transmembrane potentials of $200\ {\rm mV}$ and $500\ {\rm mV}$. We did so by applying a uniform external electric field $E_{\rm ext}$ along the z axis in the simulation box such that $V = E_{\rm ext} L_z$ where $V$ is the applied electric potential and $L_z$ the average box height \cite{Gumbart2012}, giving an applied electric field no greater than $E_{\rm ext} \approx 27.5\ {\rm mV/nm}$. 
All simulations were performed using GROMACS 4.5.5 \cite{Bekker1993,Berendsen1995,Hess2008}.
\subsection{Tension, Bending Rigidity and Fluctuation Spectrum Calculation}
We express each leaflet of the bilayer by a continuous surface expressed in terms of an expansion in complex exponentials:
\begin{align}
 \zp(x,y,z) &= \sum_{q_x} \sum_{q_y} \zpf(q_x,q_y) e^{i (x q_x + y q_y)} \\
 \zm(x,y,z) &= \sum_{q_x} \sum_{q_y} \zmf(q_x,q_y) e^{i (x q_x + y q_y)}
\end{align}
where $\zp$ ($\zm$) is the smooth surface associated with the upper (lower) layers, the $\zpf$ ($\zmf$) are the corresponding coefficients in the expansion.
 This means for the corresponding expansion coefficients of $\hf$:
\begin{equation}
\label{haverage}
 \hf(q_x,q_y) = \frac{1}{2}\left(\zpf(q_x,q_y) + \zmf(q_x,q_y)\right)
\end{equation}
The fluctuation spectrum can then be compared with Eq. \eqref{fluct}. In our simulation the tension can be measured directly by calculating the difference between the normal and the lateral pressure:
\begin{equation}
 \gamma = L_z \left( P_{zz} - \frac{1}{2}(P_{xx} + P_{yy}) \right)
\end{equation}
where $L_z$ is the box length in the $z$ direction and $P_{zz}$ is the normal and $P_{xx}$ and $P_{yy}$ are the lateral components of the averaged pressure tensor. The bending rigidity can then be inferred from the calculated fluctuation spectrum. In this paper we will assume that $\gamma = \st$, i.e., that the mechanical lateral stress acting on the membrane is the same as the tension in the fluctuation spectrum. However we point out that this issue is still controversial to this day \cite{Farago2003,Farago2004,Imparato2006,Fournier2008,Farago2011,Schmid2011}.

In contrast to previously used methods like the direct Fourier sum \cite{Brandt2011} and the binning technique \cite{Lindahl2000,Watson2011}, we obtained the fluctuation spectrum by using a least square fit with cosines and sines to the PO4 (the phosphate group) particle positions for each leaflet (see the supplementary material \cite{suppmat}). This is certainly a similar technique to the one used in \cite{Stecki2012} but no details can be found in this publication.
We note that if all the data points are evenly spaced in the simulation box, then the least square method is completely equivalent to the discrete Fourier transform applied on those points. In the range of wavenumbers we used for the calculations of the bending rigidity, the difference between the different methods is expected to be negligible. In particular we will only use the points for which $q_y = 0$ and $q<0.5\ {\rm nm^{-1}}$ (see the supplementary material for a justification\cite{suppmat}). We will postpone a discussion on this matter for a future publication. Note that choosing the PO4 particles as the definition of the layer surface is somewhat arbitrary, however we have verified that choosing another reference particle does not change the results obtained here. This is because we are looking at the average of the position between the upper and lower layer, Eq.\eqref{haverage}, and the specifics of the contributions cancel. 

\subsection{Convergence test}
For all our simulations, we checked the convergence of our fluctuation spectrum by looking at the distribution of the $| \hf(q) |^2$ after an equilibration period. Indeed, at equilibrium, a number of variables of the system will fluctuate around their mean values and have a Gaussian distribution. The $\h$ variable is one such variable of the system and therefore the distribution of the $| \hf(q) |^2$ should be exponential.
In order to quantitatively evaluate the convergence we compared the mean values of the $| \hf(q) |^2$ to their standard deviations. For an exponential distribution these two quantities are equal. The main limiting factor for the convergence of the fluctuation spectrum is the simulation time, which must be large enough in order for the largest wavelength fluctuation to sufficiently sample their phase space. The convergence criteria we chosed is that the difference between the mean and standard deviation of the fluctuations be no more than $15\%$ of the value of the mean. This criterium for convergence have been applied to all of our simulations. We believe this is the first time such a convergence test is applied to membrane fluctuations in molecular dynamics simulations and we expect it to be useful for future studies.

\section{Results and Discussion}
\subsection{Without Applied Electrical Potential}

\begin{figure}
\includegraphics{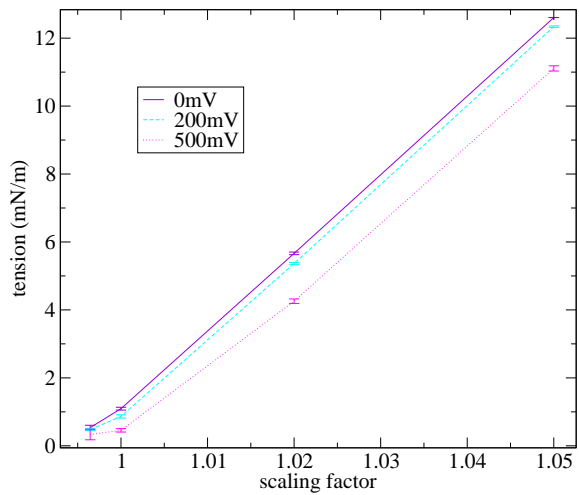}
\caption{\label{fig:tens_alpha}The tension as a function of the scaling factor with and without an applied electric potential. Note that the point for $0.996$ appears to be in a regime where a substantial part of the area is taken from the fluctuations and the membrane is not stretched.}
\end{figure}

\begin{table*}
 
\caption{\label{tab:res}Measured tension, bending rigidity and thickness for the simulations. The difference between sim1 and sim2 is the starting set of velocities. Simulation time are show in the third and fourth columns. Error bar for the thickness are standard deviation over the whole trajectory. Given error for the bending rigidity are standard deviation over block of 20 times the observed correlation time of the largest wavelength mode of the fluctuation spectrum. And error bars for the tension are standard deviations over blocks of 20 ns.}

\begin{tabular}{cc|cc|ccc}
 scaling & V (mV) & sim (ns) & sim2 (ns) & $d\ {\rm(nm)}$ & $\kappa\ (10^{-20}{\rm J})$ & $\gamma\ ({\rm mN/m})$ \\
 \hline
  & 0 &400 & 400 & $4.36\pm0.03$ & $10.4\pm 0.4$ & $0.534\pm 0.06$ \\
 0.996 & 200 & 373 & 400 & $4.36\pm0.03$  & $10.4\pm 0.9$ & $0.480   \pm 0.1$\\
  & 500 & 372  & 400 & $4.36\pm0.03$  & $10.4\pm 0.3$ & $0.33\pm 0.13$ \\
  \hline
    & 0 &200 & 200 & $4.35\pm0.02$ &$9.7\pm 1.3$ & $1.093\pm 0.08$ \\
 1.00 & 200 & 400 & 400 & $4.35\pm0.02$ &$10.8 \pm   3$ & $0.865 \pm   0.09$\\
  & 500 & 400  & 400 & $4.34\pm0.03$ &$9.9 \pm 0.8$ & $0.455\pm 0.1$ \\
  \hline
    & 0 &200 & 200 & $4.26 \pm 0.01$ &$8.9\pm   1.9$ & $5.664  \pm  0.04$ \\
 1.02 & 200 & 400 & 376 & $4.26\pm0.02$ &$9.2  \pm 1.4$ & $5.366 \pm   0.09$\\
  & 500 & 400  & 175 & $4.26\pm0.02$ &$9.1 \pm  1.2$ & $4.257  \pm  0.09$ \\
  \hline
    & 0 &200 & 195 & $4.11\pm0.01$ &$8.3 \pm  1.6$ & $12.6 \pm  0.05$ \\
 1.05 & 200 & 200 & 200 & $4.11\pm0.01$ &$8.2 \pm  1.5$ & $12.3 \pm 0.09$\\
  & 500 & 200  & 200 & $4.11\pm0.01$ &$8.0 \pm 1$ & $11.108\pm 0.08$ 
 
\end{tabular}

\end{table*}

In Table \ref{tab:res} we show the thickness, bending rigidity and mechanical tension we measured in our different simulations. As the membrane is stretched, the tension increases.
 In Figure \ref{fig:tens_alpha} we plot the tension as a function of the projected area (in our case as a function of the stretching factor as we only stretch in one direction) for all our simulations. When one increases the projected area of a fluctuating membrane, one will first pull out the excess area from the fluctuations and only at a later point will one actually start to stretch the membrane. In the region where the membrane is stretched one can recover the area expansion modulus $K_a$ \citep{Evans1990,Waheed2009}. We have only used the points for $1.00$, $1.02$ and $1.05$ in order to calculate $K_a$. We obtained $K_a \sim  224.8\pm7 {\rm mN/m}$ which is quite comparable to the experimental value $213\pm5\ {\rm mN/m}$ \citep{Henriksen2006}.
 
 \begin{figure}
      \includegraphics{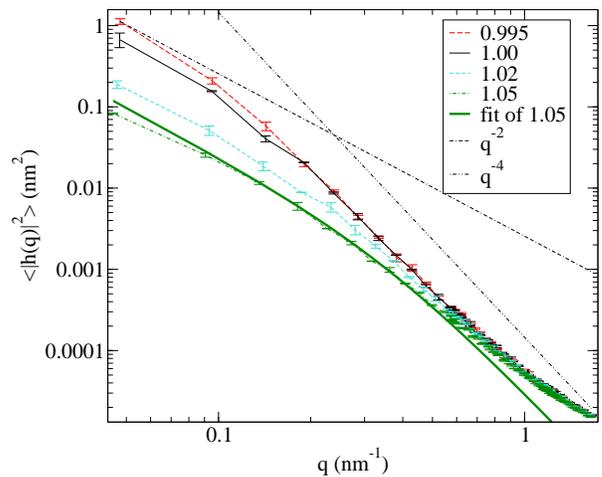}
      \caption{\label{fig:fs_no_pot}The fluctuation spectra of the simulated membranes without an applied electric potential. The $x$ scaling factor is 0.996 for the long dashed line, 1.00 for the continuous line, 1.02 for the dashed line and 1.05 for the dot-dashed line. An example of our fit used to determine the bending rigidity for the 1.05 case is shown as a thick line, note that the fit is for $q<0.5\ {\rm nm^{-1}}$. We also show lines proportional to $q^{-2}$ and $q^{-4}$ as dot-dot-dashed and dash-dash-dotted lines as guides for the eyes. Error bars are standard deviation over bins of 200 ns.}
      
\end{figure}

We show the fluctuation spectra for our simulations without an applied electric potential in Fig. \ref{fig:fs_no_pot}. Note that the overall amplitude of the fluctuation decreases as the membrane is stretched, indicating that the excess area of the membrane decreases, see Eq. \eqref{excessarea}.
According to Eq. \eqref{fluct} the fluctuation spectrum will essentially show two regimes depending on the value of $q$: for $q\ll \sqrt{\st/\kappa}$ the fluctuation spectrum becomes independent of the bending rigidity and is proportional to $1/(\st q^2)$, for $q\gg \sqrt{\st/\kappa}$ it reaches a regime independent of the tension and proportional to $1/(\kappa q^4)$ and for $q \sim \sqrt{\st/\kappa}$ the fluctuation should have a behavior between a $q^{-2}$ and a $q^{-4}$ line. We can see on the figure that in the $q^{-2}$ regime of the fluctuation spectra are decreasing as the scaling factor is increased. This is coherent with the increase of the tension we observe  in Table \ref{tab:res}. We can then obtain the bending rigidity by fitting the fluctuation spectra. Using the mechanical tension $\gamma$ for $\sigma_0$, box side lengths and temperature as measured in our simulations, the bending rigidity is our only fitting parameter in Eq.\eqref{fluct}. The results are presented in Table \ref{tab:res}. Within the error bars the bending rigidity for the $0.996$ and $1.00$ case are similar but it is clearly decreased for the $1.02$ and $1.05$ case. The decrease of the bending rigidity can be related to a decrease in the thickness of the membrane (see \citep{Evans2000} for a discussion). The value we obtained for the $1.00$ case is $\kappa\approx 10.5\times 10^{-20}\ \mathrm{J}$ and can be compared to an experimental value $ 15.8\times 10^{-20}\ \mathrm{J}$ \cite{Henriksen2006} at $25\ {\rm ^\circ C}$ without salt. Note that both the increase in temperature and the addition of salt is expected to decrease the bending rigidity \cite{Pan2008,Claessens2004} so the agreement is expected to be better for the bending rigidity for a high temperature membrane ($325\ {\rm K}$) with a large salt concentration ($\sim 100 {\rm mM}$). We are unaware of another measurement for the bending rigidity of POPC membrane using molecular dynamics simulations. We can however compare the value we obtained to the value obtained using the same (Martini) force field in \cite{Brandt2011,Watson2012}. They found  $\kappa\approx 15\times 10^{-20}\ \mathrm{J}$ for a DPPC and DMPC membrane. Only one double bond in the carbon chain is not expected to make a big difference for the bending rigidity \cite{Evans2000}.
\subsection{With Applied Electrical Potential}

\begin{figure}
      \includegraphics{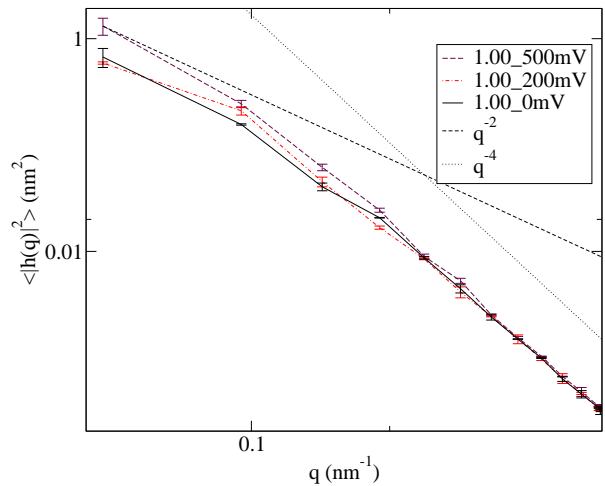}
      \caption{\label{fig:fs_1.00}Fluctuation spectra for the 1.00 case for different applied electric potentials: 0 mV (continuous line), 200 mV (long dashed line) and 500 mV (dot-dashed line). The values shown are averages from all our simulations and the error bars the standard deviation between them.}
      
\end{figure}

Next we apply an electric potential difference across our membranes. We use two values for the electric potential, 200mV and 500mV, and we measure the tension, thickness and by fitting the fluctuation spectrum we get the bending rigidity (Table \ref{tab:res}). Generally an applied electric field tend to squeeze the membrane and hence decrease the tension, as expected from theory \cite{Ambjornsson2007,Lacoste2009,Weaver1996}. We plot the fluctuation spectrum we obtained for the 1.00 case for 0, 200 and $500\ {\rm mV}$ in Figure \ref{fig:fs_1.00}. The amplitude of the large wavelength fluctuation should increase because the tension decreases. This is clearly the case with the $500\ {\rm mV}$ applied electric potential, but it is not so clear with the $200\ {\rm mV}$ applied electric potential as the first mode is found to have a lower amplitude than the $0\ {\rm mV}$ in some cases. Also, we observed no net increase or decrease of the bending rigidity with the applied electric potential and we were unable to observe a change in thickness. We conclude that, within error bars, the bending rigidity is not affected by the applied electric potential for the salt concentration we use in the simulations. Theory \cite{Ambjornsson2007,Lacoste2009} predict a contribution much lower than $k_B T$ in accordance with our observation.
Finally a note on the observed thickness. We do not observe any thickness change when we apply the electric potential, see Table \ref{tab:res}. A quick calculation, assuming that the membrane conserves its volume, shows that the thickness change should be proportional to $1-(1 + \alpha_1)/(1 + \alpha_2)$ if the membrane area changes from $\alpha_1$ to $\alpha_2$. In our case this induces a change in thickness which is less than $0.2\%$ and is bellow the error bars in Table \ref{tab:res}. Electro-compressive stresses translate into a membrane tension decrease but it is unclear from our data if there is a thickness change. This will depend on the elastic properties of the membrane as a 3 dimensional medium.

\subsection{Tension Moderation}

\begin{figure*}
      \includegraphics{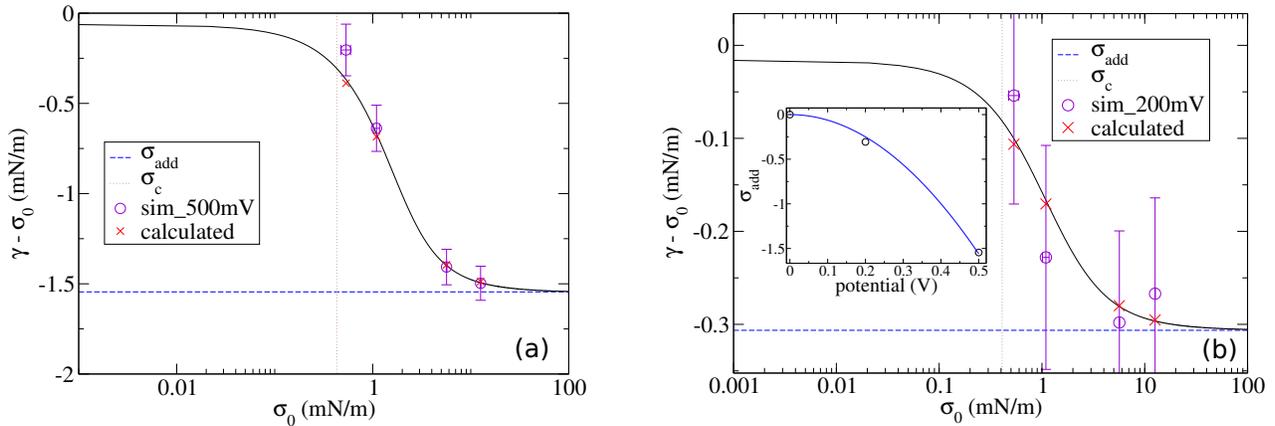}
      \caption{\label{fig:tens_mod}Additional tension due to the applied electric potential of 500 mV (a) and 200 mV (b). We plot both the measured values and their error bars (circle) as well as the calculated ones (cross) using the shown value of $\sadd$ (dashed line). The dotted line shows the value of $\sigma_c$ discussed in the text. The continuous line has been drawn by using the average value of the bending rigidity and the box size in Eq. \eqref{snoapprox} and is included as guide for the eyes. Inset: the additional tension as function of the applied potential (circle) appears to follow a quadratic law (continuous line) as given in Eq. \eqref{anzat}.}
      
\end{figure*}

Our main result is that the tension contribution from the applied electric potential is moderated depending on the initial tension of the membrane. We calculate the difference between the measured tension with an applied electric potential, $\gamma$, and the initial tension with zero electric potential, $\sigma_0$, for all our different scaling factors. Using Eq. \eqref{snoapprox} together with Eq. \eqref{fluct} we then find the value for $\sadd$ that minimizes the squared error with the values obtained in our simulations. In Figure \ref{fig:tens_mod} we show the values extracted from our simulations as well as the fitted values of $\sadd$. We obtained $\sadd = -0.31\ {\rm mN/m}$ for $200\ {\rm mV}$ and $\sadd = -1.55\ {\rm mN/m}$ for $500\ {\rm mV}$. As can be seen on the figure the contribution to the tension vanishes when $\sigma_0$ is below a critical tension $\sigma_c$ and takes its full effect only for a sufficiently stretched membrane. According to \cite{Michael2011} $\sigma_c$ can be estimated as $\sigma_c \approx K_a k_{\rm B} T /(8 \pi \kappa)$. Taking the average of the bending rigidity values in Table \ref{tab:res}, $\kappa \approx 9.75\times 10^{-20}\ \mathrm{J}$, we get $\sigma_c \approx 0.43\ {\rm mN/m}$. In the inset of Figure \ref{fig:tens_mod} we fitted the values of $\sadd$ to Eq. \eqref{anzat}. We obtained $ \mu \approx -6.2 \ {\rm mN/m/V^2}$.

Furthermore, using the equation for the tension with a large salt concentration given in \cite{Ambjornsson2007}, one can estimate $\mu\approx -\varepsilon/2d$ where $\varepsilon$ is the dielectric constant in the membrane and $2d$ the thickness of the membrane. One could have guessed this equation by assuming that the membrane behaves like a capacitor of dielectric constant $\varepsilon$ and of thickness $2d$. Using $\varepsilon \approx 2.5\varepsilon_0$, where $\varepsilon_0$ is the vacuum permittivity, and $2d \approx 4.4\ {\rm nm}$ we get $\mu \approx -5.0 \ {\rm mN/m/V^2}$, in fair agreement with our obtained values. 

\begin{figure}
      \includegraphics{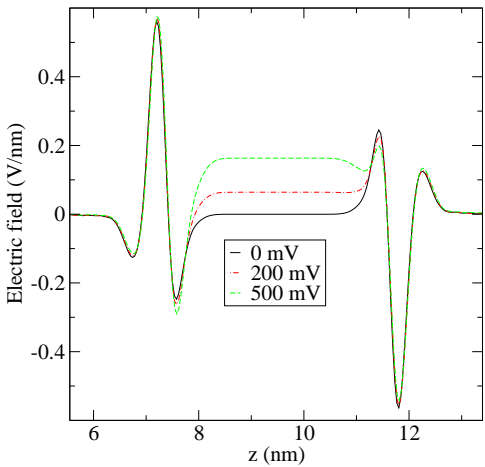}
      \caption{\label{fig:El_1.00}The electric field across the membrane in our 1.00 case with 0 mV (continuous), 200 mV (dot-dashed) and 500 mV (dashed) applied potential. Note the increase in the electric field in the region of the carbon tails of the lipids between 8 and 9 nm.}
      
\end{figure}

Next we calculate $\sadd$ directly by identifying it with minus the Maxwell stresses created by the applied electric potential. In order to evaluate this contribution we can calculate the electric field given by the average charge distribution over the z axis. Assuming that the electric field is uniform on the $x$ and $y$ direction, the Maxwell stress can be calculated as:
\begin{equation}
 T_{\rm M,zz}= -T_{\rm M,xx} = -T_{\rm M,yy} = \varepsilon_0 \frac{1}{2}\left(E_{\rm norm} \right)^2
\end{equation}
where $E_{\rm norm}$ is the integral of the average of the charge density in the $x$ and $y$ direction $\left< \rho \right>_{x,y}$:
\begin{equation}
 E_{\rm norm}(z) = \frac{1}{\varepsilon_0} \int_0^z \dd \tilde{z}\ \left< \rho \right>_{x,y}(\tilde{z})
\end{equation}
We show the electric field obtained that way in figure Fig.\ref{fig:El_1.00} for our 1.00 simullations.
The electrostatic contribution to the tension can then be evaluated as:
\begin{align}
\label{sel}
 \sel &= - \int \dd z\ \left(T_{\rm M,zz} -\frac{1}{2}\left( T_{\rm M,xx} + T_{\rm M,yy} \right) \right) \\
      &= - \varepsilon_0 \int \dd z\ \left(E_{\rm norm} \right)^2
\end{align}
where the integral is over the $z$ direction and $T_{\rm M,zz}$, $T_{\rm M,xx}$ and $T_{\rm M,yy}$ are respectively the normal and lateral Maxwell stresses. The additional contribution due to the applied potential can then be evaluated as:
\begin{equation}
\label{saddMaxwell}
 \sadd(V) = \sel(V) - \sel(0)
\end{equation}
 where $\sel(V)$ is the electrostatic tension calculated from Eq. \eqref{sel} with an applied electric potential $V$. In Table \ref{tab:Maxwell} we calculated $\sel$ and $\sadd$. We can see that for a given applied electric potential the absolute value of the different contributions of the applied electric potential to the stresses for the different scaling are the same within $10^{-2}\ {\rm mN/m}$. Furthermore we obtained $\sadd \approx -0.27 \pm 0.01 \ {\rm mN/m}$ in the $200\ {\rm mV}$ case and $\sadd \approx -1.72 \pm 0.01 \ {\rm mN/m}$ in the $500\ {\rm mV}$ case which compare well to the values we obtained from the fits of \eqref{snoapprox}, see Figure \ref{fig:tens_mod}. Also the electrostatic contribution found for the tension is on the order of (minus) the electrostatic stress found in an electroporation simulation ($1\approx2\ mN/m$ for a potential range of $0.5\approx1\ V$) \cite{Tarek2005}. Note however that further comparison are difficult because the membrane is porated is this simulation. The overall agreement between our calculated values suggests that the applied electrostatic stress does not change with the initial state of the membrane and that the moderation of the tension is a consequence of the elastic response of the membrane, due to all the non electrostatic interactions.

 \begin{table}
 
\caption{\label{tab:Maxwell}Calculated electrostatic tension in ${\rm mN/m}$. The first three columns are the values of $\sel$ as calculated from Eq.\eqref{sel}. The last two columns are the deduced values of $\sadd$, the second and third columns minus the first one (see Eq.\eqref{saddMaxwell}).}

\begin{tabular}{c|ccccc}
 streching & $0\ {\rm mV}$ & $ 200\ {\rm mV}$ & $500\ {\rm mV}$ & $\Delta 200\ {\rm mV}$ & $\Delta 500\ {\rm mV}$ \\
\hline
0.996 & -3.866 & -4.1334 & -5.584 & -0.2674 & -1.718 \\
1.00 & -3.8322 & -4.1176 & -5.5638 & -0.2854 & -1.7316 \\
1.02 & -3.7146 & -3.992 & -5.4394 & -0.2774 & -1.7248 \\
1.05 & -3.5518 & -3.81 & -5.2574 & -0.2582 & -1.7248 \\
\hline
 &  &  & mean & -0.2721 & -1.72 \\
 &  &  & std & $0.012$ & $0.011$

\end{tabular}

\end{table}

\section{Conclusion}
In this paper we have investigated the effect of an applied electrostatic potential on the mechanical properties of a membrane bilayer. We have shown that for the same applied potential the tension depends on the initial tension state of the membrane with no applied potential. The moderation happens while the electrostatic stresses are not changing significantly for different membrane stretch with the same applied potential even though the thickness of the membrane change. The lipids rearrange in order to partially accommodate for the additional stress by slightly changing their area per lipid in the membrane giving an apparent moderation of the tension. We found no significant effect of the applied electrostatic potential on the bending rigidity supposedly because of our high ion concentration.

The moderation of the tension could have significant effects on membranes and proteins functions. A floppy membrane, with large excess area, being essentially insensitive to a change of the applied potential while the potential will have its full effect on a stretched membrane, with little excess area. The tension moderation would provide a way to trigger the sensitivity of the membrane to the applied electrostatic potential, triggering the activation of mechano sensitive proteins. To strengthen that point we will note here that we observed the shift between the moderated tension and the non moderated tension for initial tension between $0.1\ {\rm mN/m}$ and $10\ {\rm mN/m}$ which is a physiologically relevant regime. For example the mechanosensitive channel MscL is activated by tension on the order of $5\ {\rm mN/m}$ \cite{Moe2005}. Also tension have been shown to influence cell motility \cite{Sheetz1996} and endocytosis \cite{Dai1997} and this will be affected by our finding.

As we showed that the electro-compressive stress effect on the membrane depends on the available excess area our finding might be of importance for electroporation phenomenon. By affecting the area per lipid the compressive stress that we discuss in this paper will certainly play a role in the probability of forming the water fingers that precede the pore formation (the so called hydrophobic pore) \cite{Tieleman2004,Tokman2013}. However the complex rearrangement of water and lipid molecules in the pore is beyond the continuous elastic sheet model used here and further investigation as to how the membrane fluctuation influence electroporation is required.

Lastly note that the external stress we discussed in this paper could be applied by other means than through an applied electrostatic potential, for example, active proteins activated by ATP could create stress in the membrane due to their conformation change possibly providing another way for the membrane to control its tension \cite{FarisPRL}.  

\section{Acknowledgements}

We thank both the Danish Center for Scientific Computing (DCSC) and the Nordic High Performance Computing (NHPC, on Gardar) for computing resources. This research was funded by Lundbeckfonden.
This research have been published in J. Chem. Phys. 139, 164902 (2013).



\bibliography{Arxiv.bib}

\begin{thebibliography}{53}%
\makeatletter
\providecommand \@ifxundefined [1]{%
 \@ifx{#1\undefined}
}%
\providecommand \@ifnum [1]{%
 \ifnum #1\expandafter \@firstoftwo
 \else \expandafter \@secondoftwo
 \fi
}%
\providecommand \@ifx [1]{%
 \ifx #1\expandafter \@firstoftwo
 \else \expandafter \@secondoftwo
 \fi
}%
\providecommand \natexlab [1]{#1}%
\providecommand \enquote  [1]{``#1''}%
\providecommand \bibnamefont  [1]{#1}%
\providecommand \bibfnamefont [1]{#1}%
\providecommand \citenamefont [1]{#1}%
\providecommand \href@noop [0]{\@secondoftwo}%
\providecommand \href [0]{\begingroup \@sanitize@url \@href}%
\providecommand \@href[1]{\@@startlink{#1}\@@href}%
\providecommand \@@href[1]{\endgroup#1\@@endlink}%
\providecommand \@sanitize@url [0]{\catcode `\\12\catcode `\$12\catcode
  `\&12\catcode `\#12\catcode `\^12\catcode `\_12\catcode `\%12\relax}%
\providecommand \@@startlink[1]{}%
\providecommand \@@endlink[0]{}%
\providecommand \url  [0]{\begingroup\@sanitize@url \@url }%
\providecommand \@url [1]{\endgroup\@href {#1}{\urlprefix }}%
\providecommand \urlprefix  [0]{URL }%
\providecommand \Eprint [0]{\href }%
\providecommand \doibase [0]{http://dx.doi.org/}%
\providecommand \selectlanguage [0]{\@gobble}%
\providecommand \bibinfo  [0]{\@secondoftwo}%
\providecommand \bibfield  [0]{\@secondoftwo}%
\providecommand \translation [1]{[#1]}%
\providecommand \BibitemOpen [0]{}%
\providecommand \bibitemStop [0]{}%
\providecommand \bibitemNoStop [0]{.\EOS\space}%
\providecommand \EOS [0]{\spacefactor3000\relax}%
\providecommand \BibitemShut  [1]{\csname bibitem#1\endcsname}%
\let\auto@bib@innerbib\@empty
\bibitem [{\citenamefont {Lee}(2004)}]{Lee2004}%
  \BibitemOpen
  \bibfield  {author} {\bibinfo {author} {\bibfnamefont {A.~G.}\ \bibnamefont
  {Lee}},\ }\href@noop {} {\bibfield  {journal} {\bibinfo  {journal} {Biochim.
  Biophys. Acta}\ }\textbf {\bibinfo {volume} {1666}},\ \bibinfo {pages} {62}
  (\bibinfo {year} {2004})}\BibitemShut {NoStop}%
\bibitem [{\citenamefont {Phillips}\ \emph {et~al.}(2009)\citenamefont
  {Phillips}, \citenamefont {Ursell}, \citenamefont {Wiggins},\ and\
  \citenamefont {Sens}}]{Phillips2009}%
  \BibitemOpen
  \bibfield  {author} {\bibinfo {author} {\bibfnamefont {R.}~\bibnamefont
  {Phillips}}, \bibinfo {author} {\bibfnamefont {T.}~\bibnamefont {Ursell}},
  \bibinfo {author} {\bibfnamefont {P.}~\bibnamefont {Wiggins}}, \ and\
  \bibinfo {author} {\bibfnamefont {P.}~\bibnamefont {Sens}},\ }\href@noop {}
  {\bibfield  {journal} {\bibinfo  {journal} {Nature}\ }\textbf {\bibinfo
  {volume} {459}},\ \bibinfo {pages} {379} (\bibinfo {year}
  {2009})}\BibitemShut {NoStop}%
\bibitem [{\citenamefont {Ambj\"ornsson}\ \emph {et~al.}(2007)\citenamefont
  {Ambj\"ornsson}, \citenamefont {Lomholt},\ and\ \citenamefont
  {Hansen}}]{Ambjornsson2007}%
  \BibitemOpen
  \bibfield  {author} {\bibinfo {author} {\bibfnamefont {T.}~\bibnamefont
  {Ambj\"ornsson}}, \bibinfo {author} {\bibfnamefont {M.~A.}\ \bibnamefont
  {Lomholt}}, \ and\ \bibinfo {author} {\bibfnamefont {P.~L.}\ \bibnamefont
  {Hansen}},\ }\href@noop {} {\bibfield  {journal} {\bibinfo  {journal} {Phys.
  Rev. E}\ }\textbf {\bibinfo {volume} {75}},\ \bibinfo {pages} {051916}
  (\bibinfo {year} {2007})}\BibitemShut {NoStop}%
\bibitem [{\citenamefont {Gauthier}\ \emph {et~al.}(2012)\citenamefont
  {Gauthier}, \citenamefont {Masters},\ and\ \citenamefont
  {Sheetz}}]{Gauthier2012}%
  \BibitemOpen
  \bibfield  {author} {\bibinfo {author} {\bibfnamefont {N.~C.}\ \bibnamefont
  {Gauthier}}, \bibinfo {author} {\bibfnamefont {T.~A.}\ \bibnamefont
  {Masters}}, \ and\ \bibinfo {author} {\bibfnamefont {M.~P.}\ \bibnamefont
  {Sheetz}},\ }\href@noop {} {\bibfield  {journal} {\bibinfo  {journal} {Trends
  Cell Bio.}\ }\textbf {\bibinfo {volume} {22}},\ \bibinfo {pages} {527}
  (\bibinfo {year} {2012})}\BibitemShut {NoStop}%
\bibitem [{\citenamefont {Loubet}\ \emph {et~al.}(2012)\citenamefont {Loubet},
  \citenamefont {Seifert},\ and\ \citenamefont {Lomholt}}]{Loubet2012}%
  \BibitemOpen
  \bibfield  {author} {\bibinfo {author} {\bibfnamefont {B.}~\bibnamefont
  {Loubet}}, \bibinfo {author} {\bibfnamefont {U.}~\bibnamefont {Seifert}}, \
  and\ \bibinfo {author} {\bibfnamefont {M.~A.}\ \bibnamefont {Lomholt}},\
  }\href@noop {} {\bibfield  {journal} {\bibinfo  {journal} {Phys. Rev. E}\
  }\textbf {\bibinfo {volume} {85}},\ \bibinfo {pages} {031913} (\bibinfo
  {year} {2012})}\BibitemShut {NoStop}%
\bibitem [{\citenamefont {Evans}\ and\ \citenamefont
  {Parsegian}(1986)}]{Evans1986}%
  \BibitemOpen
  \bibfield  {author} {\bibinfo {author} {\bibfnamefont {E.~A.}\ \bibnamefont
  {Evans}}\ and\ \bibinfo {author} {\bibfnamefont {V.~A.}\ \bibnamefont
  {Parsegian}},\ }\href@noop {} {\bibfield  {journal} {\bibinfo  {journal}
  {Proc. Natl. Acad. Sci.}\ }\textbf {\bibinfo {volume} {83}},\ \bibinfo
  {pages} {7132} (\bibinfo {year} {1986})}\BibitemShut {NoStop}%
\bibitem [{\citenamefont {Reister-Gottfried}\ \emph {et~al.}(2008)\citenamefont
  {Reister-Gottfried}, \citenamefont {Sengupta}, \citenamefont {Lorz},
  \citenamefont {Sackmann}, \citenamefont {Seifert},\ and\ \citenamefont
  {Smith}}]{Reister2008}%
  \BibitemOpen
  \bibfield  {author} {\bibinfo {author} {\bibfnamefont {E.}~\bibnamefont
  {Reister-Gottfried}}, \bibinfo {author} {\bibfnamefont {K.}~\bibnamefont
  {Sengupta}}, \bibinfo {author} {\bibfnamefont {B.}~\bibnamefont {Lorz}},
  \bibinfo {author} {\bibfnamefont {E.}~\bibnamefont {Sackmann}}, \bibinfo
  {author} {\bibfnamefont {U.}~\bibnamefont {Seifert}}, \ and\ \bibinfo
  {author} {\bibfnamefont {A.~S.}\ \bibnamefont {Smith}},\ }\href@noop {}
  {\bibfield  {journal} {\bibinfo  {journal} {Phys. Rev. Lett.}\ }\textbf
  {\bibinfo {volume} {101}},\ \bibinfo {pages} {208103} (\bibinfo {year}
  {2008})}\BibitemShut {NoStop}%
\bibitem [{\citenamefont {Faris}\ \emph {et~al.}(2009)\citenamefont {Faris},
  \citenamefont {Lacoste}, \citenamefont {Pecreaux}, \citenamefont {Joanny},
  \citenamefont {Prost},\ and\ \citenamefont {Bassereau}}]{FarisPRL}%
  \BibitemOpen
  \bibfield  {author} {\bibinfo {author} {\bibfnamefont {M.~D. E.~A.}\
  \bibnamefont {Faris}}, \bibinfo {author} {\bibfnamefont {D.}~\bibnamefont
  {Lacoste}}, \bibinfo {author} {\bibfnamefont {J.}~\bibnamefont {Pecreaux}},
  \bibinfo {author} {\bibfnamefont {J.-F.}\ \bibnamefont {Joanny}}, \bibinfo
  {author} {\bibfnamefont {J.}~\bibnamefont {Prost}}, \ and\ \bibinfo {author}
  {\bibfnamefont {P.}~\bibnamefont {Bassereau}},\ }\href@noop {} {\bibfield
  {journal} {\bibinfo  {journal} {Phys. Rev. Lett.}\ }\textbf {\bibinfo
  {volume} {102}},\ \bibinfo {pages} {038102} (\bibinfo {year}
  {2009})}\BibitemShut {NoStop}%
\bibitem [{\citenamefont {Bouvrais}\ \emph {et~al.}(2012)\citenamefont
  {Bouvrais}, \citenamefont {Cornelius}, \citenamefont {Ipsen},\ and\
  \citenamefont {Mouritsen}}]{Bouvrais2012}%
  \BibitemOpen
  \bibfield  {author} {\bibinfo {author} {\bibfnamefont {H.}~\bibnamefont
  {Bouvrais}}, \bibinfo {author} {\bibfnamefont {F.}~\bibnamefont {Cornelius}},
  \bibinfo {author} {\bibfnamefont {J.~H.}\ \bibnamefont {Ipsen}}, \ and\
  \bibinfo {author} {\bibfnamefont {O.~G.}\ \bibnamefont {Mouritsen}},\
  }\href@noop {} {\bibfield  {journal} {\bibinfo  {journal} {PNAS}\ }\textbf
  {\bibinfo {volume} {109}},\ \bibinfo {pages} {18442–18446} (\bibinfo {year}
  {2012})}\BibitemShut {NoStop}%
\bibitem [{\citenamefont {Lacoste}\ \emph {et~al.}(2009)\citenamefont
  {Lacoste}, \citenamefont {Menon}, \citenamefont {Bazant},\ and\ \citenamefont
  {Joanny}}]{Lacoste2009}%
  \BibitemOpen
  \bibfield  {author} {\bibinfo {author} {\bibfnamefont {D.}~\bibnamefont
  {Lacoste}}, \bibinfo {author} {\bibfnamefont {G.~I.}\ \bibnamefont {Menon}},
  \bibinfo {author} {\bibfnamefont {M.~Z.}\ \bibnamefont {Bazant}}, \ and\
  \bibinfo {author} {\bibfnamefont {J.-F.}\ \bibnamefont {Joanny}},\
  }\href@noop {} {\bibfield  {journal} {\bibinfo  {journal} {Eur. Phys. J. E}\
  }\textbf {\bibinfo {volume} {28}},\ \bibinfo {pages} {243} (\bibinfo {year}
  {2009})}\BibitemShut {NoStop}%
\bibitem [{\citenamefont {Lomholt}\ \emph {et~al.}(2011)\citenamefont
  {Lomholt}, \citenamefont {Loubet},\ and\ \citenamefont
  {Ipsen}}]{Michael2011}%
  \BibitemOpen
  \bibfield  {author} {\bibinfo {author} {\bibfnamefont {M.~A.}\ \bibnamefont
  {Lomholt}}, \bibinfo {author} {\bibfnamefont {B.}~\bibnamefont {Loubet}}, \
  and\ \bibinfo {author} {\bibfnamefont {J.~H.}\ \bibnamefont {Ipsen}},\
  }\href@noop {} {\bibfield  {journal} {\bibinfo  {journal} {Phys. Rev. E}\
  }\textbf {\bibinfo {volume} {83}},\ \bibinfo {pages} {011913} (\bibinfo
  {year} {2011})}\BibitemShut {NoStop}%
\bibitem [{\citenamefont {Helfrich}(1973)}]{Helfrich1973}%
  \BibitemOpen
  \bibfield  {author} {\bibinfo {author} {\bibfnamefont {W.}~\bibnamefont
  {Helfrich}},\ }\href@noop {} {\bibfield  {journal} {\bibinfo  {journal} {Z.
  Naturforsch.}\ }\textbf {\bibinfo {volume} {28c}},\ \bibinfo {pages} {693}
  (\bibinfo {year} {1973})}\BibitemShut {NoStop}%
\bibitem [{\citenamefont {Marrink}\ \emph {et~al.}(2007)\citenamefont
  {Marrink}, \citenamefont {Risselada}, \citenamefont {Yefimov}, \citenamefont
  {Tieleman},\ and\ \citenamefont {de~Vries}}]{Marrink2007}%
  \BibitemOpen
  \bibfield  {author} {\bibinfo {author} {\bibfnamefont {S.}~\bibnamefont
  {Marrink}}, \bibinfo {author} {\bibfnamefont {H.}~\bibnamefont {Risselada}},
  \bibinfo {author} {\bibfnamefont {S.}~\bibnamefont {Yefimov}}, \bibinfo
  {author} {\bibfnamefont {D.}~\bibnamefont {Tieleman}}, \ and\ \bibinfo
  {author} {\bibfnamefont {A.}~\bibnamefont {de~Vries}},\ }\href@noop {}
  {\bibfield  {journal} {\bibinfo  {journal} {J. Phys. Chem. B}\ }\textbf
  {\bibinfo {volume} {111}},\ \bibinfo {pages} {7812–7824} (\bibinfo {year}
  {2007})}\BibitemShut {NoStop}%
\bibitem [{\citenamefont {Yesylevskyy}\ \emph {et~al.}(2010)\citenamefont
  {Yesylevskyy}, \citenamefont {Schafer}, \citenamefont {Sengupta},\ and\
  \citenamefont {Marrink}}]{Yesylevskyy2010}%
  \BibitemOpen
  \bibfield  {author} {\bibinfo {author} {\bibfnamefont {S.}~\bibnamefont
  {Yesylevskyy}}, \bibinfo {author} {\bibfnamefont {L.}~\bibnamefont
  {Schafer}}, \bibinfo {author} {\bibfnamefont {D.}~\bibnamefont {Sengupta}}, \
  and\ \bibinfo {author} {\bibfnamefont {S.}~\bibnamefont {Marrink}},\
  }\href@noop {} {\bibfield  {journal} {\bibinfo  {journal} {PLoS Comput Biol}\
  }\textbf {\bibinfo {volume} {6(6)}},\ \bibinfo {pages} {1000810} (\bibinfo
  {year} {2010})}\BibitemShut {NoStop}%
\bibitem [{\citenamefont {Debye}\ and\ \citenamefont
  {H\"uckel}(1923)}]{Debye1923}%
  \BibitemOpen
  \bibfield  {author} {\bibinfo {author} {\bibfnamefont {P.}~\bibnamefont
  {Debye}}\ and\ \bibinfo {author} {\bibfnamefont {E.}~\bibnamefont
  {H\"uckel}},\ }\href@noop {} {\bibfield  {journal} {\bibinfo  {journal}
  {Phyzik Z.}\ }\textbf {\bibinfo {volume} {24}},\ \bibinfo {pages} {185}
  (\bibinfo {year} {1923})}\BibitemShut {NoStop}%
\bibitem [{\citenamefont {Brockman}(1994)}]{Brockman1994}%
  \BibitemOpen
  \bibfield  {author} {\bibinfo {author} {\bibfnamefont {H.}~\bibnamefont
  {Brockman}},\ }\href@noop {} {\bibfield  {journal} {\bibinfo  {journal}
  {Chem. Phys. Lipids}\ }\textbf {\bibinfo {volume} {73}},\ \bibinfo {pages}
  {57} (\bibinfo {year} {1994})}\BibitemShut {NoStop}%
\bibitem [{\citenamefont {Yang}\ \emph {et~al.}(2008)\citenamefont {Yang},
  \citenamefont {Mayer}, \citenamefont {Wickremasinghe},\ and\ \citenamefont
  {Hafner}}]{Yang2008}%
  \BibitemOpen
  \bibfield  {author} {\bibinfo {author} {\bibfnamefont {Y.}~\bibnamefont
  {Yang}}, \bibinfo {author} {\bibfnamefont {K.~M.}\ \bibnamefont {Mayer}},
  \bibinfo {author} {\bibfnamefont {N.~S.}\ \bibnamefont {Wickremasinghe}}, \
  and\ \bibinfo {author} {\bibfnamefont {J.~H.}\ \bibnamefont {Hafner}},\
  }\href@noop {} {\bibfield  {journal} {\bibinfo  {journal} {Biophys. J}\
  }\textbf {\bibinfo {volume} {95}},\ \bibinfo {pages} {5193} (\bibinfo {year}
  {2008})}\BibitemShut {NoStop}%
\bibitem [{\citenamefont {Berendsen}\ \emph {et~al.}(1984)\citenamefont
  {Berendsen}, \citenamefont {Postma}, \citenamefont {DiNola},\ and\
  \citenamefont {Haak}}]{Berendsen1984}%
  \BibitemOpen
  \bibfield  {author} {\bibinfo {author} {\bibfnamefont {H.~J.~C.}\
  \bibnamefont {Berendsen}}, \bibinfo {author} {\bibfnamefont {J.~P.~M.}\
  \bibnamefont {Postma}}, \bibinfo {author} {\bibfnamefont {A.}~\bibnamefont
  {DiNola}}, \ and\ \bibinfo {author} {\bibfnamefont {J.~R.}\ \bibnamefont
  {Haak}},\ }\href@noop {} {\bibfield  {journal} {\bibinfo  {journal} {J. Chem.
  Phys.}\ }\textbf {\bibinfo {volume} {81}},\ \bibinfo {pages} {3684–3690}
  (\bibinfo {year} {1984})}\BibitemShut {NoStop}%
\bibitem [{\citenamefont {Nose}(1984)}]{Nose1984}%
  \BibitemOpen
  \bibfield  {author} {\bibinfo {author} {\bibfnamefont {S.~A.}\ \bibnamefont
  {Nose}},\ }\href@noop {} {\bibfield  {journal} {\bibinfo  {journal} {Mol.
  Phys.}\ }\textbf {\bibinfo {volume} {52}},\ \bibinfo {pages} {255–268}
  (\bibinfo {year} {1984})}\BibitemShut {NoStop}%
\bibitem [{\citenamefont {Hoover}(1985)}]{Hoover1985}%
  \BibitemOpen
  \bibfield  {author} {\bibinfo {author} {\bibfnamefont {W.~G.}\ \bibnamefont
  {Hoover}},\ }\href@noop {} {\bibfield  {journal} {\bibinfo  {journal} {Phys.
  Rev. A}\ }\textbf {\bibinfo {volume} {31}},\ \bibinfo {pages} {1695–1697}
  (\bibinfo {year} {1985})}\BibitemShut {NoStop}%
\bibitem [{\citenamefont {Parrinello}\ and\ \citenamefont
  {Rahman}(1981)}]{Parrinello1981}%
  \BibitemOpen
  \bibfield  {author} {\bibinfo {author} {\bibfnamefont {M.}~\bibnamefont
  {Parrinello}}\ and\ \bibinfo {author} {\bibfnamefont {A.}~\bibnamefont
  {Rahman}},\ }\href@noop {} {\bibfield  {journal} {\bibinfo  {journal} {J.
  Appl. Phys.}\ }\textbf {\bibinfo {volume} {52}},\ \bibinfo {pages}
  {7182–7190} (\bibinfo {year} {1981})}\BibitemShut {NoStop}%
\bibitem [{\citenamefont {Nose}\ and\ \citenamefont {Klein}(1983)}]{Nose1983}%
  \BibitemOpen
  \bibfield  {author} {\bibinfo {author} {\bibfnamefont {S.}~\bibnamefont
  {Nose}}\ and\ \bibinfo {author} {\bibfnamefont {M.~L.}\ \bibnamefont
  {Klein}},\ }\href@noop {} {\bibfield  {journal} {\bibinfo  {journal} {Mol.
  Phys.}\ }\textbf {\bibinfo {volume} {50}},\ \bibinfo {pages} {1055–1076}
  (\bibinfo {year} {1983})}\BibitemShut {NoStop}%
\bibitem [{\citenamefont {Darden}\ \emph {et~al.}(1993)\citenamefont {Darden},
  \citenamefont {York},\ and\ \citenamefont {Pedersend}}]{Darden1993}%
  \BibitemOpen
  \bibfield  {author} {\bibinfo {author} {\bibfnamefont {T.}~\bibnamefont
  {Darden}}, \bibinfo {author} {\bibfnamefont {D.}~\bibnamefont {York}}, \ and\
  \bibinfo {author} {\bibfnamefont {L.}~\bibnamefont {Pedersend}},\ }\href@noop
  {} {\bibfield  {journal} {\bibinfo  {journal} {J. Chem. Phys.}\ }\textbf
  {\bibinfo {volume} {98}},\ \bibinfo {pages} {10089–10092} (\bibinfo {year}
  {1993})}\BibitemShut {NoStop}%
\bibitem [{\citenamefont {Essmann}\ \emph {et~al.}(1995)\citenamefont
  {Essmann}, \citenamefont {Perera}, \citenamefont {Berkowitz}, \citenamefont
  {Darden}, \citenamefont {Lee},\ and\ \citenamefont {Pedersen}}]{Essmann1995}%
  \BibitemOpen
  \bibfield  {author} {\bibinfo {author} {\bibfnamefont {U.}~\bibnamefont
  {Essmann}}, \bibinfo {author} {\bibfnamefont {L.}~\bibnamefont {Perera}},
  \bibinfo {author} {\bibfnamefont {M.~L.}\ \bibnamefont {Berkowitz}}, \bibinfo
  {author} {\bibfnamefont {T.}~\bibnamefont {Darden}}, \bibinfo {author}
  {\bibfnamefont {H.}~\bibnamefont {Lee}}, \ and\ \bibinfo {author}
  {\bibfnamefont {L.~G.}\ \bibnamefont {Pedersen}},\ }\href@noop {} {\bibfield
  {journal} {\bibinfo  {journal} {J. Chem. Phys.}\ }\textbf {\bibinfo {volume}
  {103}},\ \bibinfo {pages} {8577–8592} (\bibinfo {year} {1995})}\BibitemShut
  {NoStop}%
\bibitem [{\citenamefont {Gumbart}\ \emph {et~al.}(2012)\citenamefont
  {Gumbart}, \citenamefont {Khalili-Araghi}, \citenamefont {Sotomayor},\ and\
  \citenamefont {Roux}}]{Gumbart2012}%
  \BibitemOpen
  \bibfield  {author} {\bibinfo {author} {\bibfnamefont {J.}~\bibnamefont
  {Gumbart}}, \bibinfo {author} {\bibfnamefont {F.}~\bibnamefont
  {Khalili-Araghi}}, \bibinfo {author} {\bibfnamefont {M.}~\bibnamefont
  {Sotomayor}}, \ and\ \bibinfo {author} {\bibfnamefont {B.}~\bibnamefont
  {Roux}},\ }\href@noop {} {\bibfield  {journal} {\bibinfo  {journal}
  {BBA-Biomembranes}\ }\textbf {\bibinfo {volume} {1818}},\ \bibinfo {pages}
  {294} (\bibinfo {year} {2012})}\BibitemShut {NoStop}%
\bibitem [{\citenamefont {Bekker}\ \emph {et~al.}(1993)\citenamefont {Bekker},
  \citenamefont {Berendsen}, \citenamefont {Dijkstra}, \citenamefont
  {Achterop}, \citenamefont {van Drunen}, \citenamefont {van~der Spoel},
  \citenamefont {Sijbers}, \citenamefont {Keegstra}, \citenamefont {Reitsma},\
  and\ \citenamefont {Renardus}}]{Bekker1993}%
  \BibitemOpen
  \bibfield  {author} {\bibinfo {author} {\bibfnamefont {H.}~\bibnamefont
  {Bekker}}, \bibinfo {author} {\bibfnamefont {H.~J.~C.}\ \bibnamefont
  {Berendsen}}, \bibinfo {author} {\bibfnamefont {E.~J.}\ \bibnamefont
  {Dijkstra}}, \bibinfo {author} {\bibfnamefont {S.}~\bibnamefont {Achterop}},
  \bibinfo {author} {\bibfnamefont {R.}~\bibnamefont {van Drunen}}, \bibinfo
  {author} {\bibfnamefont {D.}~\bibnamefont {van~der Spoel}}, \bibinfo {author}
  {\bibfnamefont {A.}~\bibnamefont {Sijbers}}, \bibinfo {author} {\bibfnamefont
  {H.}~\bibnamefont {Keegstra}}, \bibinfo {author} {\bibfnamefont
  {B.}~\bibnamefont {Reitsma}}, \ and\ \bibinfo {author} {\bibfnamefont
  {M.~K.~R.}\ \bibnamefont {Renardus}},\ }\href@noop {} {\emph {\bibinfo
  {title} {Physics Computing 92: Gromacs: A parallel computer for molecular
  dynamics simulations}}}\ (\bibinfo  {publisher} {R A De Groot J Nadrchal},\
  \bibinfo {year} {1993})\BibitemShut {NoStop}%
\bibitem [{\citenamefont {Berendsen}\ \emph {et~al.}(1995)\citenamefont
  {Berendsen}, \citenamefont {van~der Spoel},\ and\ \citenamefont {van
  Drunen}}]{Berendsen1995}%
  \BibitemOpen
  \bibfield  {author} {\bibinfo {author} {\bibfnamefont {H.~J.~C.}\
  \bibnamefont {Berendsen}}, \bibinfo {author} {\bibfnamefont {D.}~\bibnamefont
  {van~der Spoel}}, \ and\ \bibinfo {author} {\bibfnamefont {R.}~\bibnamefont
  {van Drunen}},\ }\href@noop {} {\bibfield  {journal} {\bibinfo  {journal}
  {Comp. Phys. Comm.}\ }\textbf {\bibinfo {volume} {91}},\ \bibinfo {pages}
  {43} (\bibinfo {year} {1995})}\BibitemShut {NoStop}%
\bibitem [{\citenamefont {Hess}\ \emph {et~al.}(2008)\citenamefont {Hess},
  \citenamefont {Kutzner}, \citenamefont {van~der Spoel},\ and\ \citenamefont
  {Lindahl}}]{Hess2008}%
  \BibitemOpen
  \bibfield  {author} {\bibinfo {author} {\bibfnamefont {B.}~\bibnamefont
  {Hess}}, \bibinfo {author} {\bibfnamefont {C.}~\bibnamefont {Kutzner}},
  \bibinfo {author} {\bibfnamefont {D.}~\bibnamefont {van~der Spoel}}, \ and\
  \bibinfo {author} {\bibfnamefont {E.}~\bibnamefont {Lindahl}},\ }\href@noop
  {} {\bibfield  {journal} {\bibinfo  {journal} {J. Chem. Theory Comp.}\
  }\textbf {\bibinfo {volume} {4(3)}},\ \bibinfo {pages} {435} (\bibinfo {year}
  {2008})}\BibitemShut {NoStop}%
\bibitem [{\citenamefont {Farago}\ and\ \citenamefont
  {Pincus}(2003)}]{Farago2003}%
  \BibitemOpen
  \bibfield  {author} {\bibinfo {author} {\bibfnamefont {O.}~\bibnamefont
  {Farago}}\ and\ \bibinfo {author} {\bibfnamefont {P.}~\bibnamefont
  {Pincus}},\ }\href@noop {} {\bibfield  {journal} {\bibinfo  {journal} {Eur.
  Phys. J.}\ }\textbf {\bibinfo {volume} {11}},\ \bibinfo {pages} {399}
  (\bibinfo {year} {2003})}\BibitemShut {NoStop}%
\bibitem [{\citenamefont {Farago}\ and\ \citenamefont
  {Pincus}(2004)}]{Farago2004}%
  \BibitemOpen
  \bibfield  {author} {\bibinfo {author} {\bibfnamefont {O.}~\bibnamefont
  {Farago}}\ and\ \bibinfo {author} {\bibfnamefont {P.}~\bibnamefont
  {Pincus}},\ }\href@noop {} {\bibfield  {journal} {\bibinfo  {journal} {J.
  Chem. Phys.}\ }\textbf {\bibinfo {volume} {120}},\ \bibinfo {pages} {2934}
  (\bibinfo {year} {2004})}\BibitemShut {NoStop}%
\bibitem [{\citenamefont {Imparato}(2006)}]{Imparato2006}%
  \BibitemOpen
  \bibfield  {author} {\bibinfo {author} {\bibfnamefont {A.}~\bibnamefont
  {Imparato}},\ }\href@noop {} {\bibfield  {journal} {\bibinfo  {journal} {J.
  Chem. Phys.}\ }\textbf {\bibinfo {volume} {124}},\ \bibinfo {pages} {154714}
  (\bibinfo {year} {2006})}\BibitemShut {NoStop}%
\bibitem [{\citenamefont {Fournier}\ and\ \citenamefont
  {Barbetta}(2008)}]{Fournier2008}%
  \BibitemOpen
  \bibfield  {author} {\bibinfo {author} {\bibfnamefont {J.-B.}\ \bibnamefont
  {Fournier}}\ and\ \bibinfo {author} {\bibfnamefont {C.}~\bibnamefont
  {Barbetta}},\ }\href@noop {} {\bibfield  {journal} {\bibinfo  {journal}
  {Phys. Rev. Lett.}\ }\textbf {\bibinfo {volume} {100}},\ \bibinfo {pages}
  {078103} (\bibinfo {year} {2008})}\BibitemShut {NoStop}%
\bibitem [{\citenamefont {Farago}(2011)}]{Farago2011}%
  \BibitemOpen
  \bibfield  {author} {\bibinfo {author} {\bibfnamefont {O.}~\bibnamefont
  {Farago}},\ }\href@noop {} {\bibfield  {journal} {\bibinfo  {journal} {Phys.
  Rev. E}\ }\textbf {\bibinfo {volume} {84}},\ \bibinfo {pages} {051914}
  (\bibinfo {year} {2011})}\BibitemShut {NoStop}%
\bibitem [{\citenamefont {Schmid}(2011)}]{Schmid2011}%
  \BibitemOpen
  \bibfield  {author} {\bibinfo {author} {\bibfnamefont {F.}~\bibnamefont
  {Schmid}},\ }\href@noop {} {\bibfield  {journal} {\bibinfo  {journal} {EPL}\
  }\textbf {\bibinfo {volume} {95}},\ \bibinfo {pages} {28008} (\bibinfo {year}
  {2011})}\BibitemShut {NoStop}%
\bibitem [{\citenamefont {Brandt}\ \emph {et~al.}(2011)\citenamefont {Brandt},
  \citenamefont {Braun}, \citenamefont {Sachs}, \citenamefont {Nagle},\ and\
  \citenamefont {Edholm}}]{Brandt2011}%
  \BibitemOpen
  \bibfield  {author} {\bibinfo {author} {\bibfnamefont {E.~G.}\ \bibnamefont
  {Brandt}}, \bibinfo {author} {\bibfnamefont {A.~R.}\ \bibnamefont {Braun}},
  \bibinfo {author} {\bibfnamefont {J.~N.}\ \bibnamefont {Sachs}}, \bibinfo
  {author} {\bibfnamefont {J.~F.}\ \bibnamefont {Nagle}}, \ and\ \bibinfo
  {author} {\bibfnamefont {O.}~\bibnamefont {Edholm}},\ }\href@noop {}
  {\bibfield  {journal} {\bibinfo  {journal} {Biophys. J.}\ }\textbf {\bibinfo
  {volume} {100}},\ \bibinfo {pages} {2104–2111} (\bibinfo {year}
  {2011})}\BibitemShut {NoStop}%
\bibitem [{\citenamefont {Lindahl}\ and\ \citenamefont
  {Edholm}(2000)}]{Lindahl2000}%
  \BibitemOpen
  \bibfield  {author} {\bibinfo {author} {\bibfnamefont {E.}~\bibnamefont
  {Lindahl}}\ and\ \bibinfo {author} {\bibfnamefont {O.}~\bibnamefont
  {Edholm}},\ }\href@noop {} {\bibfield  {journal} {\bibinfo  {journal}
  {Biophys. J.}\ }\textbf {\bibinfo {volume} {79}},\ \bibinfo {pages} {426}
  (\bibinfo {year} {2000})}\BibitemShut {NoStop}%
\bibitem [{\citenamefont {Watson}\ \emph {et~al.}(2011)\citenamefont {Watson},
  \citenamefont {Penev}, \citenamefont {Welch},\ and\ \citenamefont
  {Brown}}]{Watson2011}%
  \BibitemOpen
  \bibfield  {author} {\bibinfo {author} {\bibfnamefont {M.~C.}\ \bibnamefont
  {Watson}}, \bibinfo {author} {\bibfnamefont {E.~S.}\ \bibnamefont {Penev}},
  \bibinfo {author} {\bibfnamefont {P.~M.}\ \bibnamefont {Welch}}, \ and\
  \bibinfo {author} {\bibfnamefont {F.~L.~H.}\ \bibnamefont {Brown}},\
  }\href@noop {} {\bibfield  {journal} {\bibinfo  {journal} {J. Chem. Phys.}\
  }\textbf {\bibinfo {volume} {135}},\ \bibinfo {pages} {244701} (\bibinfo
  {year} {2011})}\BibitemShut {NoStop}%
\bibitem [{sup()}]{suppmat}%
  \BibitemOpen
  \href@noop {} {}\bibinfo {note} {See supplementary material at [URL will be
  inserted by AIP] for details on the technique used.}\BibitemShut {Stop}%
\bibitem [{\citenamefont {Stecki}(2012)}]{Stecki2012}%
  \BibitemOpen
  \bibfield  {author} {\bibinfo {author} {\bibfnamefont {J.}~\bibnamefont
  {Stecki}},\ }\href@noop {} {\bibfield  {journal} {\bibinfo  {journal} {J.
  Chem. Phys.}\ }\textbf {\bibinfo {volume} {137}},\ \bibinfo {pages} {116102}
  (\bibinfo {year} {2012})}\BibitemShut {NoStop}%
\bibitem [{\citenamefont {Evans}\ and\ \citenamefont
  {Rawicz}(1990)}]{Evans1990}%
  \BibitemOpen
  \bibfield  {author} {\bibinfo {author} {\bibfnamefont {E.}~\bibnamefont
  {Evans}}\ and\ \bibinfo {author} {\bibfnamefont {W.}~\bibnamefont {Rawicz}},\
  }\href {\doibase {10.1103/PhysRevLett.64.2094}} {\bibfield  {journal}
  {\bibinfo  {journal} {Physical Review Letters}\ }\textbf {\bibinfo {volume}
  {64}},\ \bibinfo {pages} {2094} (\bibinfo {year} {1990})}\BibitemShut
  {NoStop}%
\bibitem [{\citenamefont {Waheed}\ and\ \citenamefont
  {Edholm}(2009)}]{Waheed2009}%
  \BibitemOpen
  \bibfield  {author} {\bibinfo {author} {\bibfnamefont {Q.}~\bibnamefont
  {Waheed}}\ and\ \bibinfo {author} {\bibfnamefont {O.}~\bibnamefont
  {Edholm}},\ }\href@noop {} {\bibfield  {journal} {\bibinfo  {journal}
  {Biophys. J.}\ }\textbf {\bibinfo {volume} {97}},\ \bibinfo {pages} {2754}
  (\bibinfo {year} {2009})}\BibitemShut {NoStop}%
\bibitem [{\citenamefont {Henriksen}\ \emph {et~al.}(2006)\citenamefont
  {Henriksen}, \citenamefont {Rowat}, \citenamefont {Brief}, \citenamefont
  {Hsueh}, \citenamefont {Thewalt}, \citenamefont {Zuckermann},\ and\
  \citenamefont {Ipsen}}]{Henriksen2006}%
  \BibitemOpen
  \bibfield  {author} {\bibinfo {author} {\bibfnamefont {J.}~\bibnamefont
  {Henriksen}}, \bibinfo {author} {\bibfnamefont {A.~C.}\ \bibnamefont
  {Rowat}}, \bibinfo {author} {\bibfnamefont {E.}~\bibnamefont {Brief}},
  \bibinfo {author} {\bibfnamefont {Y.~W.}\ \bibnamefont {Hsueh}}, \bibinfo
  {author} {\bibfnamefont {J.~L.}\ \bibnamefont {Thewalt}}, \bibinfo {author}
  {\bibfnamefont {M.~J.}\ \bibnamefont {Zuckermann}}, \ and\ \bibinfo {author}
  {\bibfnamefont {J.~H.}\ \bibnamefont {Ipsen}},\ }\href@noop {} {\bibfield
  {journal} {\bibinfo  {journal} {Biophys. J.}\ }\textbf {\bibinfo {volume}
  {90}},\ \bibinfo {pages} {1639} (\bibinfo {year} {2006})}\BibitemShut
  {NoStop}%
\bibitem [{\citenamefont {Rawicz}\ \emph {et~al.}(2000)\citenamefont {Rawicz},
  \citenamefont {Olbrich}, \citenamefont {McIntosh}, \citenamefont {Needham},\
  and\ \citenamefont {Evans}}]{Evans2000}%
  \BibitemOpen
  \bibfield  {author} {\bibinfo {author} {\bibfnamefont {W.}~\bibnamefont
  {Rawicz}}, \bibinfo {author} {\bibfnamefont {K.~C.}\ \bibnamefont {Olbrich}},
  \bibinfo {author} {\bibfnamefont {T.}~\bibnamefont {McIntosh}}, \bibinfo
  {author} {\bibfnamefont {D.}~\bibnamefont {Needham}}, \ and\ \bibinfo
  {author} {\bibfnamefont {E.}~\bibnamefont {Evans}},\ }\href@noop {}
  {\bibfield  {journal} {\bibinfo  {journal} {Biophysical Journal}\ }\textbf
  {\bibinfo {volume} {79}},\ \bibinfo {pages} {328} (\bibinfo {year}
  {2000})}\BibitemShut {NoStop}%
\bibitem [{\citenamefont {Pan}\ \emph {et~al.}(2008)\citenamefont {Pan},
  \citenamefont {Tristram-Nagle}, \citenamefont {Kuc\v{e}rka},\ and\
  \citenamefont {Nagle}}]{Pan2008}%
  \BibitemOpen
  \bibfield  {author} {\bibinfo {author} {\bibfnamefont {J.}~\bibnamefont
  {Pan}}, \bibinfo {author} {\bibfnamefont {S.}~\bibnamefont {Tristram-Nagle}},
  \bibinfo {author} {\bibfnamefont {N.}~\bibnamefont {Kuc\v{e}rka}}, \ and\
  \bibinfo {author} {\bibfnamefont {J.~F.}\ \bibnamefont {Nagle}},\ }\href@noop
  {} {\bibfield  {journal} {\bibinfo  {journal} {Biophys. J.}\ }\textbf
  {\bibinfo {volume} {94}},\ \bibinfo {pages} {117} (\bibinfo {year}
  {2008})}\BibitemShut {NoStop}%
\bibitem [{\citenamefont {Claessens}\ \emph {et~al.}(2004)\citenamefont
  {Claessens}, \citenamefont {van Oort}, \citenamefont {Leermakers},
  \citenamefont {Hoekstra},\ and\ \citenamefont {Stuart}}]{Claessens2004}%
  \BibitemOpen
  \bibfield  {author} {\bibinfo {author} {\bibfnamefont {M.~M. A.~E.}\
  \bibnamefont {Claessens}}, \bibinfo {author} {\bibfnamefont {B.~F.}\
  \bibnamefont {van Oort}}, \bibinfo {author} {\bibfnamefont {F.~A.~M.}\
  \bibnamefont {Leermakers}}, \bibinfo {author} {\bibfnamefont {F.~A.}\
  \bibnamefont {Hoekstra}}, \ and\ \bibinfo {author} {\bibfnamefont {M.~A.~C.}\
  \bibnamefont {Stuart}},\ }\href@noop {} {\bibfield  {journal} {\bibinfo
  {journal} {Biophys. J.}\ }\textbf {\bibinfo {volume} {87}},\ \bibinfo {pages}
  {3882} (\bibinfo {year} {2004})}\BibitemShut {NoStop}%
\bibitem [{\citenamefont {Watson}\ \emph {et~al.}(2012)\citenamefont {Watson},
  \citenamefont {Brandt}, \citenamefont {Welch},\ and\ \citenamefont
  {Brown}}]{Watson2012}%
  \BibitemOpen
  \bibfield  {author} {\bibinfo {author} {\bibfnamefont {M.~C.}\ \bibnamefont
  {Watson}}, \bibinfo {author} {\bibfnamefont {E.~G.}\ \bibnamefont {Brandt}},
  \bibinfo {author} {\bibfnamefont {P.~M.}\ \bibnamefont {Welch}}, \ and\
  \bibinfo {author} {\bibfnamefont {F.~L.~H.}\ \bibnamefont {Brown}},\
  }\href@noop {} {\bibfield  {journal} {\bibinfo  {journal} {Phys. Rev. Lett.}\
  }\textbf {\bibinfo {volume} {109}},\ \bibinfo {pages} {028102} (\bibinfo
  {year} {2012})}\BibitemShut {NoStop}%
\bibitem [{\citenamefont {Weaver}\ and\ \citenamefont
  {Chizmadzhev}(1996)}]{Weaver1996}%
  \BibitemOpen
  \bibfield  {author} {\bibinfo {author} {\bibfnamefont {J.~C.}\ \bibnamefont
  {Weaver}}\ and\ \bibinfo {author} {\bibfnamefont {Y.}~\bibnamefont
  {Chizmadzhev}},\ }\href@noop {} {\bibfield  {journal} {\bibinfo  {journal}
  {Bioelectroch. Bioener.}\ }\textbf {\bibinfo {volume} {41}},\ \bibinfo
  {pages} {135} (\bibinfo {year} {1996})}\BibitemShut {NoStop}%
\bibitem [{\citenamefont {Tarek}(2005)}]{Tarek2005}%
  \BibitemOpen
  \bibfield  {author} {\bibinfo {author} {\bibfnamefont {M.}~\bibnamefont
  {Tarek}},\ }\href@noop {} {\bibfield  {journal} {\bibinfo  {journal}
  {Biophys. J.}\ }\textbf {\bibinfo {volume} {88}},\ \bibinfo {pages} {4045}
  (\bibinfo {year} {2005})}\BibitemShut {NoStop}%
\bibitem [{\citenamefont {Moe}\ and\ \citenamefont {Blount}(2005)}]{Moe2005}%
  \BibitemOpen
  \bibfield  {author} {\bibinfo {author} {\bibfnamefont {P.}~\bibnamefont
  {Moe}}\ and\ \bibinfo {author} {\bibfnamefont {P.}~\bibnamefont {Blount}},\
  }\href@noop {} {\bibfield  {journal} {\bibinfo  {journal} {Biochem.}\
  }\textbf {\bibinfo {volume} {44}},\ \bibinfo {pages} {12239} (\bibinfo {year}
  {2005})}\BibitemShut {NoStop}%
\bibitem [{\citenamefont {Sheetz}\ and\ \citenamefont
  {Dai.}(1996)}]{Sheetz1996}%
  \BibitemOpen
  \bibfield  {author} {\bibinfo {author} {\bibfnamefont {M.~P.}\ \bibnamefont
  {Sheetz}}\ and\ \bibinfo {author} {\bibfnamefont {J.}~\bibnamefont {Dai.}},\
  }\href@noop {} {\bibfield  {journal} {\bibinfo  {journal} {Trends Cell
  Biol.}\ }\textbf {\bibinfo {volume} {6}},\ \bibinfo {pages} {85} (\bibinfo
  {year} {1996})}\BibitemShut {NoStop}%
\bibitem [{\citenamefont {Dai}\ \emph {et~al.}(1997)\citenamefont {Dai},
  \citenamefont {Ting-Beall},\ and\ \citenamefont {Sheetz}}]{Dai1997}%
  \BibitemOpen
  \bibfield  {author} {\bibinfo {author} {\bibfnamefont {J.}~\bibnamefont
  {Dai}}, \bibinfo {author} {\bibfnamefont {H.~P.}\ \bibnamefont {Ting-Beall}},
  \ and\ \bibinfo {author} {\bibfnamefont {M.~P.}\ \bibnamefont {Sheetz}},\
  }\href@noop {} {\bibfield  {journal} {\bibinfo  {journal} {J. Gen. Physiol.}\
  }\textbf {\bibinfo {volume} {110}},\ \bibinfo {pages} {1} (\bibinfo {year}
  {1997})}\BibitemShut {NoStop}%
\bibitem [{\citenamefont {Tieleman}(2004)}]{Tieleman2004}%
  \BibitemOpen
  \bibfield  {author} {\bibinfo {author} {\bibfnamefont {D.~P.}\ \bibnamefont
  {Tieleman}},\ }\href@noop {} {\bibfield  {journal} {\bibinfo  {journal} {BMC
  Biochem.}\ }\textbf {\bibinfo {volume} {5}},\ \bibinfo {pages} {10} (\bibinfo
  {year} {2004})}\BibitemShut {NoStop}%
\bibitem [{\citenamefont {Tokman}\ \emph {et~al.}(2013)\citenamefont {Tokman},
  \citenamefont {Lee}, \citenamefont {Levine}, \citenamefont {Ho},\ and\
  \citenamefont {Colvin}}]{Tokman2013}%
  \BibitemOpen
  \bibfield  {author} {\bibinfo {author} {\bibfnamefont {M.}~\bibnamefont
  {Tokman}}, \bibinfo {author} {\bibfnamefont {J.}~\bibnamefont {Lee}},
  \bibinfo {author} {\bibfnamefont {Z.}~\bibnamefont {Levine}}, \bibinfo
  {author} {\bibfnamefont {M.-C.}\ \bibnamefont {Ho}}, \ and\ \bibinfo {author}
  {\bibfnamefont {M.}~\bibnamefont {Colvin}},\ }\href@noop {} {\bibfield
  {journal} {\bibinfo  {journal} {PLoS ONE}\ }\textbf {\bibinfo {volume} {8}},\
  \bibinfo {pages} {e61111} (\bibinfo {year} {2013})}\BibitemShut {NoStop}%
\end{thebibliography}%

\end{document}